\documentstyle[12pt]{article}
\title{The $Q^2$ Dependence of the Sum Rules for Structure\\
Functions of Polarized $e(\mu)N$ Scattering \thanks{Talk presented
at the Workshop on High Energy Polarization Phenomena,
Ringberg, Germany, February 24-28, 1997, unpublished}}
\author{B.L.Ioffe \thanks{E-mail: ioffe@vitep5.itep.ru}\\
Institute of Theoretical and Experimental Physics\\
117259, Moscow, Russia}
\date{}
\begin{document}

\maketitle

\newcommand{\be}{\begin{equation}}
\newcommand{\ee}{\end{equation}}

\def\la{\mathrel{\mathpalette\fun <}}
\def\ga{\mathrel{\mathpalette\fun >}}
\def\fun#1#2{\lower3.6pt\vbox{\baselineskip0pt\lineskip.9pt
\ialign{$\mathsurround=0pt#1\hfil##\hfil$\crcr#2\crcr\sim\crcr}}}

\centerline{\bf{Abstract}}

\vspace{5mm}
The nonperturbative $Q^2$- dependence of the sum rules for the structure
functions of polarized $e(\mu)N$ scattering is discussed. The determination
of twist-4 corrections to the structure functions at high $Q^2$ by QCD sum
rules is reviewed and critically analyzed. It is found that in the case of
the Bjorken sum rule the twist-4 correction is small at $Q^2 > 5
GeV^2$ and does not influence the value of $\alpha_s$ determined from this
sum rule. However, the accuracy of the today experimental data is
insufficient to reliably determine $\alpha_s$ from the Bjorken sum rule. For
the singlet sum rule -- $p+n$ -- the QCD sum rule gives only the order of
magnitude of twist-4 correction. At low and intermediate $Q^2$ the model is
presented which realizes a smooth connection of the Gerasimov-Drell-Hearn
sum rules at $Q^2 = 0$ with the sum rules for $\Gamma_{p,n}(Q^2)$ at high
$Q^2$. The model is in a good agreement with the experiment.

\newpage
{\large \bf 1.~ Introduction}

\vspace{3mm}
In the last few years there is a strong interest to the problem of nucleon
spin structure: how nucleon spin is distributed among its constituents -
quarks and gluons. New experimental data continuously appear and precision
increases (for recent reviews see \cite{1}, \cite{2}). One of the most
important item of the information comes from the measurements of the first
moment of the spin-dependent nucleon structure functions $g_1(x)$ which
determine the parts of nucleon spin carried by u, d and s quarks and gluons.
The level of the accuracy of the data is now such that the account of
nonperturbative $Q^2$ dependence -- the so called twist-4 terms-- is of
importance when comparing the data with the Bjorken and Ellis-Jaffe sum
rules at high $Q^2$. On the other side,  at low and intermediate $Q^2$ a
smooth connection of the sum rules for the first moments of $g_1(x)$ with
the Gerasimov-Drell-Hearn (GDH) sume rules [3,4] is theoretically expected.
This connection can be realized through nonperturbative $Q^2$-dependence
only. In this paper I discuss such nonperturbative $Q^2$-dependence of the
sum rules.

First, recall the standard definition of the structure functions for
scattering of polarized electrons or muons on polarized nucleon. They are
characterized by the imaginary part of the forward scattering amplitude of
virtual photon on the nucleon (see, e.g., \cite{5})

\be
Im~T^a_{\mu \nu}(q,p) = \frac{2\pi}{m}~\epsilon_{\mu \nu \lambda \sigma}
q_{\lambda} \Biggl [s_{\sigma} G_1(\nu, q^2)+
 \frac{1}{m^2} (s_{\sigma} \nu - (sq)p_{\sigma}) G_2(\nu, q^2)\Biggr ]
\ee
Here $\nu = pq, q^2 = q^2_0 - {\bf q}^2 < 0, q$ and $p$ are virtual photon
and nucleon momenta, $s_{\sigma}$ is nucleon spin 4-vector, $G_1(\nu, q^2)$
and $G_2(\nu, q^2)$ are two spin-dependent nucleon structure functions, $m$
is the nucleon mass. Index $a$ in eq.(1) means antisymmetrization in virtual
photon polarization indeces $\mu, \nu$. Below I will consider only the
structure function $G_1(\nu, q^2)$ related to the scaling structure function
$g_1(x, Q^2)$ by

\be
\frac{\nu}{m^2} G_1(\nu, q^2) = g_1(x, Q^2),
\ee
where $Q^2 = -q^2, x = Q^2/2 \nu$. The first moment of the structure
function $g_1$ is defined as

\be
\Gamma_{p,n}(Q^2) = \int \limits^{1}_{0}~ dx~g_{1; p,n}(x, Q^2)
\ee

The presentation of the material in the paper is divided into two parts. The
first part deals with the case of high $Q^2$. I discuss the determination
of twist-4 contributions to $\Gamma_{p,n}$ by QCD sum rules and estimate
their influence on the values of $\alpha_s(Q^2)$ as well as on the values of
$\Delta u, \Delta d, \Delta s$ -- the parts of the proton spin projection,
carried by $u,d,s$ quarks.  In the second part the case of low and
intermediate $Q^2 \la 1 GeV^2$ is considered in the framework of the model
which realizes the smooth connection of GDH sum rule at $Q^2 = 0$ with the
asymptotic form of $\Gamma_{p,n}(Q^2)$ at high $Q^2$

\vspace{5cm}
{\large \bf 2. ~High $Q^2$}

\vspace{3mm}
At high $Q^2$ with the account of twist-4 contributions $\Gamma_{p,n}(Q^2)$
have the form

$$
\Gamma_{p,n}(Q^2) = \Gamma^{as}_{p,n}(Q^2) + \Gamma^{tw 4}_{p,n}(Q^2)
$$
\be
\Gamma^{as}_{p,n}(Q^2) = \frac{1}{12} \Biggl \{ [1 - a - 3.58 a^2 - 20.2 a^3
- c a^4 ] [\pm g_A + \frac{1}{3}a_8]
\ee

\be
+ \frac{4}{3} [1 - \frac{1}{3} a - 0.55 a^2 - 4.45 a^3] \Sigma \Biggr \} -
\frac{N_f}{18 \pi} \alpha_s(Q^2) \Delta g(Q^2)
\ee

\be
\Gamma^{tw4}_{p,n}(Q^2) = \frac{b_{p,n}}{Q^2}
\ee
In eq.(5) $a = \alpha_s(Q^2)/\pi, g_A$ is the $\beta$-decay axial coupling
constant, $g_A = 1.260 \pm 0.002$ [6]

\be
g_A = \Delta u - \Delta d ~~~~ a_8 = \Delta u + \Delta d - 2 \Delta s ~~~~
\Sigma = \Delta u + \Delta d + \Delta s.
\ee
$\Delta u, \Delta d, \Delta s, \Delta g$ are parts of the nucleon
spin projections carried by $u, d, s$ quarks and gluons:

\be
\Delta q = \int \limits^{1}_{0} \Biggl [q_+(x) - q_-(x) \Biggr ]
\ee
where $q_+(x), q_-(x)$ are quark distributions with spin projection
parallel (antiparallel) to nucleon spin and a similar definition takes place
for $\Delta g$. The coefficients of perturbative series were calculated in
\cite{7,8,9,10}, the numerical values in (5) correspond to the number of
flavours $N_f = 3$, the coefficient $c$ was estimated in \cite{11}, $c
\approx 130$.  In the renormalization scheme chosen in \cite{7,8,9,10} $a_8$
and $\Sigma$ are $Q^2$-independent. In the assumption of the exact $SU(3)$
flavour symmetry of the octet axial current matrix elements over baryon
octet states $a_8 = 3F - D = 0.59 \pm 0.02$ \cite{12}.

Strictly speaking, in (5) the separation of terms proportional to $\Sigma$
and $\Delta g$ is arbitrary, since the operator product expansion (OPE) has
only one singlet in flavour twist-2 operator for the first moment of the
polarized structure function -- the operator of singlet axial current
$j^{(0)}_{\mu 5}(x) = \sum \limits_{q}~ \bar{q}_i(x) \gamma_{\mu} \gamma_5 q, ~~ q =
u,d,s$. The separation of terms proportional to $\Sigma$ and $\Delta g$ is
outside the framework of OPE and depends on the infrared cut-off. The
expression used in (5) is based on the physical assumption that the
virtualities $p^2$ of gluons in the nucleon are much larger than light quark
mass squares, $\mid p^2\mid \gg m^2_q$ \cite{13} and that the infrared
cut-off is chosen in a way providing the standard form of axial anomaly
\cite{14}. (See \cite{15} for review and details).

Let us now discuss twist-4 contributions to $\Gamma_{p,n}$ (4) and estimate
the coefficients $b_{p,n}$ in (6). The general theory of twist-4
contributions to the first moment of $g_{1; p,n}$ was formulated by Shuryak
and Vainstein \cite{16}. They have found

\newpage
$$
(\Gamma_p \pm \Gamma_n)_{twist 4} = -\frac{8}{9}~ \frac{1}{Q^2}~ C_{S, NS}
\Biggl [ \langle \langle V \rangle \rangle_{S, NS} - \frac{1}{4} \langle
\langle V \rangle \rangle_{S, NS} \Biggr ]
$$
\be
+ \frac{2}{9}~ \frac{m^2}{Q^2} \int \limits^{1}_{0} x^2 dx g_{1; p\pm n} (x)
\ee
Here the matrix elements in the double angular brackets mean the matrix
elements of the operators (for $u$-quarks)

$$
U^u_{\mu} = g \bar{u}~\tilde{G}^n_{\mu \nu} \gamma_{\nu} (\lambda^n/2) u,~~~
\tilde{G}^n_{\mu \nu} = \frac{1}{2} \epsilon_{\mu \nu \lambda \sigma}
G^n_{\lambda \sigma}
$$
\be
V^u_{\mu \nu, \sigma} = \frac{1}{2} g \bar{u} \tilde{G}^n_{\mu \nu}
\gamma_{\sigma} \frac{1}{2} \lambda^n u + (\nu \leftrightarrow \sigma)
\ee
defined as

$$
\langle N \vert U_{\mu} \vert N \rangle = s_{\mu} \langle \langle U \rangle
\rangle
$$
\be
\langle N \vert V_{\mu \nu, \sigma} \vert N \rangle = S_{\nu, \sigma}
A_{\mu, \nu} s_{\mu} p_{\nu} p_{\sigma} \langle \langle V \rangle \rangle,
\ee
where $S$ and $A$ are symmetrization and antisymmetrization operators. The
indeces $S$ (singlet) and $NS$ (nonsinglet) correspond to $S \rightarrow u +
d + (18/5)s, ~~ NS \rightarrow u-d$ and

\be
C_S = \frac{5}{18}, ~~~~ C_{NS} = \frac{1}{6}
\ee
The last term in (9) is small and  can be safely neglected.

The matrix elements $\langle \langle U \rangle \rangle_{S, NS}$ and $\langle
\langle V \rangle \rangle_{S, NS}$ were calculated by Balitsky, Braun and
Koleshichenko (BBK) \cite{17}  using the QCD sum rule method in external
field.  (The explicit form of (9) was taken from \cite{17}, where the errors
in \cite{16} were corrected - see \cite{17}, Errata). In order to perform
the calculations one should add to QCD Lagrangian the terms

\be
\delta L = U_{\mu} A_{\mu}, ~~~~ \delta L = V_{\mu \nu, \sigma} A_{\mu, \nu,
\sigma}
\ee
correspondingly, in case of $\langle N \vert U_{\mu} \vert N \rangle$ and
$\langle N \vert V_{\mu \nu, \sigma} \vert n \rangle$, calculations, where
$A_{\mu}$ and $A_{\mu \nu, \sigma}$ are constant external fields. Then the
polarization operator

\be
\Pi = i \int~ dx e^{ipx} \langle 0 \vert T \{ \eta(x),
\bar{\eta}(0)  \} \vert 0 \rangle
\ee
is considered and the terms in $\Pi(p)$ linear in external fields are
separated. In (14) $\eta(x)$ is the quark current with nucleon quantum
numbers. For proton \cite{18}

\be
\eta_p(x) = \epsilon^{abc} u^a(x) C \gamma_{\mu} u^b(x) \gamma_5
\gamma_{\mu} d^c(x),
\ee
$u^a(x), d^c(x)$ -- are $u$ and $d$- quark fields, $a, b, c$ -- are colour
indeces. An essential ingredient of the QCD sum rules method in external
field is the appearance of induced by external field vacuum condensates in
OPE \cite{19}. To determine these condensates some additional sum rules are
used.  In the case in view there are two such condensates:

\be
\langle 0 \vert U_{\mu} \vert 0 \rangle_A = P A_{\mu}, ~~~ \langle 0 \vert
V_{\mu \nu, \sigma} \vert 0 \rangle_A = R A_{\mu \nu, \sigma}
\ee
Their values were estimated by BBK \cite{17}:
$P = 3.10^{-3} (\pm 30\%) GeV^6, R = 1.10^{-3} (\pm 100\%) GeV^6$. BBK
accounted in OPE the operators up to dimension 8. The sum rules for
nonsinglet matrix elements are

$$
\langle \langle U^{NS} \rangle \rangle + A^{NS}_U M^2 = - \frac{1}{2
\tilde{\lambda^2}} e^{m^2/M^2} \Biggl \{\frac{8}{9} M^2
\frac{\alpha_s(M^2)}{\pi} \times
$$

\be
\times \int \limits^{W^2}_{0} sds~ e^{-s/M^2} \Biggl (W^2 + s~ ln
\frac{W^2}{s} \Biggr ) - \frac{1}{9} b M^4 E_1 \Biggl (\frac{W^2}{M^2} +
\Biggr)
\ee

$$
+ \frac{32}{27}~M^2 \frac{\alpha_s(M^2)}{\pi}~ a^2 (ln \frac{W^2}{M^2} +
1.03) + \frac{8}{9} \pi^2 PM^2 - \frac{2}{3} m^2_0 a^2 \Biggr \}
$$

$$
\langle \langle V^{NS} \rangle \rangle + A^{NS}_V M^2 = - \frac{1}{2
\tilde{\lambda}^2} e^{m^2/M^2} \Biggl \{- \frac{52}{135}~
\frac{\alpha_s(M^2)}{\pi} M^2 \times
$$

$$
\times \int \limits^{W^2}_0 sds~ e^{-s/M^2} (W^2 + s ln \frac{W^2}{s} ) -
\frac{2}{9}~ b M^4 E_1 \Biggl (\frac{W^2}{M^2}\Biggr ) -
$$
\be
- \frac{80}{27}~ \frac{\alpha_s(M^2)}{\pi} M^2 a^2 (ln \frac{W^2}{M^2} +
1.89) + \pi^2 RM^2 - \frac{4}{9} m^2_0 a^2 \Biggr \},
\ee
where $M^2$ is the Borel parameter

$$
\tilde{\lambda}^2 = 32 \pi^4 \lambda^2, ~~~ \lambda = \langle 0 \vert \eta
\vert N \rangle, ~~~ \tilde{\lambda}^2 = 2.1 GeV^6
$$
$$
a = -(2 \pi)^2 \langle 0 \vert \bar{q} q \vert 0 \rangle = 0.67 GeV^3
~~(\mbox{at}~~ M^2 = 1 GeV^2)
$$

$$
b = \frac{\alpha_s}{\pi} \langle 0 \vert G^2_{\mu \nu} \vert 0 \rangle (2
\pi)^2 = 0.5 GeV^4, ~~~ m^2_0 = 0.8 GeV^2
$$
\be
E_1(z) = 1 - (1 + z)e^{-z}
\ee
and $W^2$ is the continuum threshold, $W^2 = 2.3 GeV^2$.

The terms $A^{NS}_{U,V}$ in the l.h.s of the sum rules are background
contributions corresponding to nondiagonal transitions in the
phenomenological parts of the sum rules:

$$
N \rightarrow \mbox{interaction with external current} \rightarrow N^*
$$
They should be separated by studying the $W^2$ dependence of the sum rules,
e.g., by applying the differential operator $1 - M^2 \partial/\partial M^2$
to the sum rule. Unfortunately, such procedure deteriorates the accuracy of
the results.

The sum rules (17),(18) differ from the original ones found by BBK. In the
BBK sum rules the first and the third terms in r.h.s. contain the
ultraviolet cut-off. As was shown in \cite{20}, this is not correct and stems
from the fact that BBK used one-variable dispersion relation in $p^2$ for
the vertex function $\Gamma(p^2)$. Instead of $\Gamma(p^2)$ the general
vertex function $\Gamma(p^2_1, p^2_2, Q^2)$ must be considered and for this
function double dispersion relation in $p^2_1$ and $p^2_2$ with subtraction
terms must be written. After going to the limit $p^2_1 \rightarrow p^2_2, ~~
q^2 \rightarrow 0$ the correct dispersion relation is obtained. In this
double dispersion relation representation for $\Gamma(p^2, p^2, 0)$ no
ultraviolet cut-off appears, but a more strong assumption about the relation
of physical spectrum to $\Gamma(p^2, p^2, 0)$ found in QCD calculation is
needed. In expressions (17),(18) the hypothesis of local duality was
assumed when the interval \\
$0 < \mid p^2_1\mid, \mid p^2_2\mid < W^2$ in the dispersion
representation for $\Gamma(p^2, p^2)$ corresponds to the nucleon pole while
everything outside this interval -- to continuum. (See \cite{20} for
details).

The calculation of the r.h.s. of (17),(18) in the interval of the Borel
parameter \\
$0.8 < M^2 < 1.2 GeV^2$ gives for the l.h.s approximately

\be
\langle \langle U^{NS} \rangle \rangle + A^{NS}_U M^2 == 0.15 - 0.23 M^2
\ee

\be
\langle \langle V^{NS} \rangle \rangle + A^{NS}_V M^2 = 0.64 - 0.17 M^2
\ee
in $GeV^2$. For the twist-4 correction (6) we have according to (9):

\be
b_{p-n} = -\frac{1}{6} \cdot \frac{8}{9} \Biggl [\langle \langle U^{NS}
\rangle \rangle - \frac{1}{4} \langle \langle V^{NS} \rangle \rangle \Biggr
] = 0.0015
\ee
i.e., practically zero. This zero result arises due to strong compensation
of two terms in (22). That is why the analysis of uncertainties in the
calculation is necessary. A serious uncertainty in the calculation of
$\langle \langle U^{NS} \rangle \rangle$ comes from the fact that the main
contribution to the r.h.s. of (17) is given by the last term -- the
operator of dimension 8 and some doubts appear about a possible role of
higher dimension operators. Recently Oganesian had
calculated  the contribution of the dimension-10 operator to
the sum rule (17) in the framework of the factorization hypothesis
\cite{21}.  His result is:

\be
\langle \langle U^{NS} \rangle \rangle_{dim 10} = \frac{1}{9}
\frac{e^{m^2/M^2}}{\tilde{\lambda}^2} m^4_0 a^2 \Biggl (1 + \frac{m^2}{2 M^2}
\Biggr ) \approx 0.05 GeV^2
\ee
In the case of $\langle \langle V^{NS} \rangle \rangle$ the main
contribution comes from the third term in (18) - the operator of dimension
6. So, one may believe, the the higher dimension operators are not very
important here. Other possible sources of error are: 1) the large background
term, proportional to $M^2$ in (20);  a much stronger influence of the
continuum threshold $W^2$ on the sum rules (17),(18) comparing with usual
QCD sum rules, where the $W^2$ dependence appears only through correction
terms $\simeq e^{-W^2/M^2}$; 3) the role of anomalous dimensions which are
disregarded in (17),(18), but can destroy compensation in (22);\\
4) even
uncertainties in the chosen numerical values of QCD parameters ($\Lambda,
a^2$ etc.) can influence this compensation. Adding (23) to (22) and
estimating the uncertainties as one half of each term in (22) we have
finally

\be
b_{p-n} = -0.006 \pm 0.012
\ee

The calculation of the sum rules for singlet matrix elements, which I do not
present here, results in

\be
\langle \langle U^S \rangle \rangle + A^S_U M^2 = 0.082 - 0.23 M^2
\ee

\be
\langle \langle V^S \rangle \rangle + A^S_V M^2 = -0.22 + 0.05 M^2
\ee
The contribution of the dimension-10 operator to $\langle \langle U^S
\rangle \rangle$ was found to be \cite{21}

\be
\langle \langle U^S \rangle \rangle_{dim 10} = -\frac{2}{9} ~
\frac {e^{m^2/M^2}}{\tilde{\lambda}^2} m^4_0 a^2 \Biggl (1 + \frac{m^2}{2
M^2} \Biggr ) \approx -0.10
\ee
and is even larger than $\langle \langle
U^S \rangle \rangle$ from (25).  For $\langle \langle V^S \rangle \rangle$
the main contribution to the sum rule comes also from the highest dimension
8 operator.

There are also other very serious drawbacks of the twist-4
matrix elements calculations done by BBK \cite{17} in the singlet case:

1. BBK assumed that $s$-quarks do not contribute to the spin structure
functions and instead of singlet operator considered the octet one.

2. When determining the induced by external field vacuum condensates, which
are very important in the calculation of $\langle \langle U^S \rangle
\rangle$, the corresponding sum rule was saturated by $\eta$-meson, what is
wrong. (Even saturation by $\eta^{\prime}$ meson would not be correct, since
$\eta^{\prime}$ is not a Goldstone).

3. The calculation of the singlet axial current matrix element over the
nucleon state -- $\langle N \mid j^0_{\mu 5} \mid 0 \rangle = s_{\mu}
\Sigma$ -- by QCD sum rule fails: it was shown that the OPE series diverges
at the scale $M^2 \sim 1 GeV^2$ \cite{22}. It is very probably that the same
situation takes place in the calculation of $\langle \langle U^S \rangle
\rangle$ and $\langle \langle V^S \rangle \rangle$.

For all these reasons the results for twist-4 corrections, following from
(25),(26)

\be
b_{p+n} = \frac {5}{18} \cdot \frac{8}{9} \Biggl [\langle \langle U^S
 \rangle \rangle - \frac{1}{4}\langle \langle V^S \rangle \rangle \Biggr ] =
 -0.035
\ee
may be considered only as correct by the order of magnitude.

Bearing in mind the values and uncertainties of twist-4 corrections to
$\Gamma_{p,n}$ let us compare the theory with the recent experimental data
\cite{1,2}.

Table 1 shows the combined experimental data \cite{1} obtained in SMC
\cite{1} and SLAC \cite{2} experiments, transferred to $Q^2 = 5 GeV^2$ in
comparison with the theoretical values  of the Ellis-Jaffe and Bjorken sum
rule. The Bjorken sum rule was calculated according to (5), the magnitude of
the twist-4 correction was taken from (24), the $\alpha_s$ value $\alpha_s(5
GeV^2) = 0.276$, corresponding to $\alpha_s(m_z) = 0.117$ and
$\Lambda^{(3)}_{\bar{MS}}= 360 MeV$ (the latter in two loops).

\vspace{5mm}
\centerline{\underline{Table 1}}

\vspace{3mm}
\noindent
Combined experimental data [1] of SMC \cite{1} and SLAC \cite{2} for
$\Gamma_p, \Gamma_n$ and $\Gamma_p - \Gamma_n$ at $Q^2 = 5 GeV^2$ in
comparison with theoretical values of Ellis-Jaffe and Bjorken sum rules.

\vspace{3mm}
\begin{tabular}{|c|c|c|c|}\hline
& $\Gamma_p$ & $\Gamma_n$ &
              $\Gamma_p - \Gamma_n$\\ \hline Combined data & $0.142 \pm
0.011$&$-0.061 \pm 0.016$&$0.202 \pm 0.022$\\ \hline
Ellis-Jaffe/Bjorken&$0.168 \pm 0.005$&$-0.013 \pm 0.005$&$0.181 \pm 0.002$\\
\hline
\end{tabular}

\vspace{3mm}
\noindent
The $\Lambda^{(3)}_{\bar{MS}} = 200 MeV~~ (\alpha_s(5 GeV^2) = 0.215,
\alpha_s(m_z) = 0.106$) would give instead $\Gamma_p - \Gamma_n = 0.189$.
The Ellis-Jaffe sum rule prediction was calculated according to (5) where
$\Delta s = 0$, i.e. $\Sigma = a_8 = 0.59$ was put, the last - gluonic term
in (5) -- was omitted and the twist-4 contribution (28) was included into
the error.The errors in the second line of Table 1 are only from
uncertainties of twist-4 terms. A possible error arising from violation of
the $SU(3)$ flavour symmetry only weakly affects the Ellis-Jaffe prediction:
$10\%$ variation of $a_8$ results in $\Delta \Gamma_p = \Delta \Gamma_n =
0.008$. Therefore, as follows from Table 1, the Ellis-Jaffe sum rule is in a
definite contradiction with experiment - a nonzero value of $\Delta s$ is
necessary. From the experimental value of $\Gamma_p - \Gamma_n$ presented in
Table 1 and from the Bjorken sum rule

\be
\Gamma_p - \Gamma_n = \frac{g_A}{6} [1 - a - 3.58a^2 - 20.2 a^3 - 130
a^4] + \frac{b_{p-n}}{Q^2}
\ee
one can determine the coupling constant
$\alpha_s(Q^2)$ at $Q^2 = 5 GeV^2$.  The result is

\be
\alpha_s(5 GeV^2) = 0.116^{+ 0.16}_{-0.44} \pm 0.014
\ee
The first error is experimental, the second comes from the uncertainty in
the twist-4 correction (24). The value $\alpha_s(5 GeV^2)$ is nonsatisfactory
because of large errors and because of that the central point corresponds to
$\Lambda^{(3)}_{\bar{MS}} \approx 15 MeV$, what is unacceptable. A serious
reduction of experimental errors in $\Gamma_p - \Gamma_n$ -- by factors
$\simeq 3-4$ is necessary in order that  one could determine QCD coupling
constant $\alpha_s(Q^2)$ from the Bjorken sum rule with a reasonable
accuracy, say, to distinguish the cases of large and small values of
$\Lambda^{(3)} \approx 350 MeV$ and $\Lambda^{(3)} \approx 200 MeV$, which
are now under dispute.

Parts of the proton spin, carried by $u, d$ and $s$-quarks can be calculated
from the data of Table 1 and eq.'s(5-7). The results of the overall fit
presented in ref.1 are

$$
\Delta u = 0.82 \pm 0.02 ~~~~ \Delta d = -0.43 \pm 0.02, ~~~~ \Delta s =
-0.10 \pm 0.02
$$
\be
\Sigma = 0.29 \pm 0.06
\ee
However, these results are strongly dependent on the values of $\alpha_s$ in
the analysis as well as from what set of the data they are determined. For
example, at $\alpha_s(5 GeV^2) = 0.276~~ (\alpha_s(m_z) = 0.117)$ from
$\Gamma_p, \Gamma_n$ and $\Gamma_p + \Gamma_n$ it follows correspondingly
$\Sigma = 0.48; ~0.145;~ 0.246$ -- the values which are  partly outside one
standard deviation quoted in (31). A selfconsistent value  of $\Sigma
\approx 0.22$ can be found from all data -- $\Gamma_p, \Gamma_n$ and $\Gamma
+ \Gamma_n$, if we put for $\alpha_s(5 GeV^2)$ the central value (30), but
this is nonacceptable, as mentioned above. For this reason, I think, that
the errors in (31) are underestimated. (The errors from twist-4 corrections
$\delta \Sigma = \pm 0.03,~\delta \Delta q = \pm 0.01$ are not included in
(31), since they are smaller).

Up to now the contribution of gluons -- the last term in (5) -- was
disregarded. The estimation of this term can be done \cite{15}, if we assume,
that at $1 GeV^2$ the quark model is valid and in the relation of the
conservation angular momentum

\be
\frac{1}{2}\Sigma + \Delta g(Q^2) + L_z(Q^2) = \frac{1}{2}
\ee
the orbital momentum term $L_z(Q^2)$  can be neglected. Assuming $\Sigma(1
GeV^2) \approx 0.4$, we find $\Delta g(1GeV^2)=0.3$.
Then from the evolution equation \cite{23} (see also \cite{15}), it follows
that $\Delta g (5 GeV^2)\approx 0.6$, what results in increasing of $\Sigma$
by $\delta \Sigma \approx 0.08$ and in corresponding increasing of $\Delta
u,~ \Delta d, ~\Delta s$  by $\delta q = \delta \Sigma/3 \approx 0.03$.

\vspace{5mm}
{\large \bf 3.~ Low and Intermediate $Q^2$.}

\vspace{3mm}
{\large \bf ~~~~Connection with Gerasimov-Drell-Hearn Sum Rule.}

\vspace{3mm}
 Consider the forward scattering amplitude of polarized real photon on
 polarized nucleon. The spin dependent part of the amplitude is expressed
 through one invariant function. In the lab.system we can write

\be
e^{(2)}_i T^a_{ik} e^{(1)}_k = i~\frac{\nu}{m^2}~\varepsilon_{ikl}e^{(2)}_i
e^{(1)}_k s_l ~S_1(\nu,0),
\ee
where $e^{(1)}_k,~ e^{(2)}_k~(i,k = 1,2,3)$ are the polarization vectors of
initial and final photons, $s_l$ -- is the vector of nucleon spin, $\nu = pq
= m \omega$ in the l.s.  The second, equal to zero argument of the invariant
function $S_1(\nu, Q^2)$ means, that the photon is real, $Q^2 = 0$.
Comparing  (33)  with the general expression (1) one can easily see, that

\be
Im~S_1(\nu, 0) = 2\pi~G_1(\nu,0)
\ee
(the factor in front of the function $G_2(\nu, q^2)$  in (1) vanishes for
the case of the real photon).

As follows from Regge theory, the leading Regge pole trajectory, determining
the high energy behavior of $S_1(\nu,0)$ is the trajectory of $a_1$  Regge
pole \cite{24,5}

\be
S_1(\nu,0)_{\nu\to \infty} \sim \nu^{\alpha_{a_1}(0) - 1},
\ee
where $\alpha_{a_1}$ is the intercept of $a_1$ trajectory. The value of
$\alpha_{a_1}(0)$  is not completely certain $\alpha_{a_1}(0) \approx -(0.3
\div 0.0)$, but definitely it is negative. Therefore, $S_1(\nu,0)_{\nu \to
\infty} < 1/\nu$  and unsubtracted dispersion relation can be written for it
\cite{5}:

\be
S_1 (\nu, 0) = 4~\int\limits^{\infty}_0~
\nu^{\prime}d\nu^{\prime}\frac{G_1(\nu^{\prime}, 0)}{\nu^{\prime 2} - \nu^2}
\ee
Consider now $S_1(\nu, 0)$ in the limit $\nu \to 0$. According to F.Low
theorem $S_1(\nu, 0)$ is expressed through the static nucleon properties and
the direct calculation gives

\be
S_1 (\nu, 0)_{\nu \to 0} = -\kappa^2,
\ee
where $\kappa$ is the nucleon anomalous magnetic moment, $\kappa_p = 1.79,
~\kappa_n = -1.91$. The substitution of (37)  into (36) gives
the Gerasimov-Drell-Hearn (GDH) \cite{3,4} sum rule

\be
\int\limits^{\infty}_0~\frac{d\nu}{\nu}~G_1(\nu,0) = -\frac{1}{4}\kappa^2
\ee

An important remark: the forward spin dependent photon-nucleon scattering
amplitude has no nucleon pole in case of real photon. This means that there
is no nucleon contribution in the l.h.s. of (38) -- all contributions come
from excited states. The GDH sum rule is very nontrivial!

Till now there are no direct experimental checks of GDH sum rule, since the
experiments on photoproduction by polarized photon on polarized proton (or
neutron)  are absent. What was done \cite{25,26,27} -- is the indirect check,
when the parameters of nucleon resonances, determined in nonpolarized
photo-- or electroproduction were substituted into the l.h.s. of (38).  In
this way with resonances up to $W = 1.8$GeV it was obtained \cite{25,26,27}
(the numbers below are taken from \cite{28}):

\be
~~~~~~~~~~~~~~~~~~~~\mbox{l.h.s.~ ~of~~(38) }~~~~~~~~~~\mbox{r.h.s.~~of
~~(38)}
\ee
$$
\mbox{proton} ~~~~~~~~~~~~~ -1.03 ~~~~~~~~~~~~~~~~~~~~-0.8035
$$
$$ \mbox{neutron}~~~~~~~~~~~-0.83 ~~~~~~~~~~~~~~~~~~~~-0.9149$$
The errors in (39)  are such, that with the account of only resonances up to
$W = 1.8$GeV the l.h.s. and the r.h.s. of (38) are not in agreement -- a
nonresonant contribution is needed.

In order to connect the GDH sum rule with $\Gamma_{p,n}(Q^2)$ consider the
integrals \cite{5}

\be
I_{p,n}(Q^2) =
\int\limits^{\infty}_{Q^2/2}~\frac{d\nu}{\nu}G_{_1;p,n}(\nu,Q^2)
\ee
Using (2)  and changing the integration variable $\nu$ to $x$, (40) can be
also identically written as

\be
I_{p,n}(Q^2) = \frac{2 m^2}{Q^2}~\int\limits^1_0~dx~ g_{1;p,n} (x,Q^2) =
\frac{2m^2}{Q^2}\Gamma_{p,n} (Q^2)
\ee
At $Q^2 = 0$

\be
I_p(0) = -\frac{1}{4}\kappa^2_p = - 0.8035;~~~I_n (0) =
-\frac{1}{4}\kappa^2_n = =0.9149; ~~~I_p(0) - I_n(0) = 0.1114
\ee
and $\Gamma_{p,n} = 0$. Sometimes it is convenient to express $\Gamma(Q^2)$
in terms of the electroproduction cross sections $\sigma_{1/2}(\nu,Q^2)$ and
$\sigma_{3/2}(\nu,Q^2)$, corresponding to the projections $1/2$  and $1/3$
of the total  photon-nucleon spin upon the photon momentum direction, as
well as the quantity $\sigma_I(\nu,Q^2),$  describing the interference of
transverse and longitudinal virtual photon  polarizations \cite{5}:

$$\Gamma(Q^2) = \frac{Q^2}{16 \pi^2\alpha}~\int \frac{d\nu}{\nu}~\frac{1 -
x}{1 + (m^2Q^2/\nu^2)}~\Biggl [ \sigma_{1/2}(\nu, Q^2) -
\sigma_{3/2}(\nu,Q^2) +$$

\be
+ 2 \frac{Q m}{\nu}~\sigma_I(\nu, Q^2)\Biggr ]
\ee
(The normalization of the cross sections is chosen in  a way, that the
virtual photon flux is assumed to be equal to that of real photons with the
energy fixed by  condition, that the masses of hadronic states produced
by the real and virtual photons are equal.)  $\sigma_I(\nu,Q^2)$ satisfies
the inequality

\be
\sigma_I < \sqrt{R}\sigma_T, ~~~\sigma_T = \sigma_{1/2} + \sigma_{3/2}, ~~~R
= \sigma_L/\sigma_T
\ee
and practically the last term in (43)  is small and as a rule can be
neglected.

The schematic $Q^2$  dependence of $I_p(Q^2),~I_n(Q^2)$  and $I_p(Q^2) -
I_n(Q^2)$  is plotted in Fig.1. The case of $I_p(Q^2)$ is especially
interesting: $I_p(Q^2)$  is positive, small and decreasing at $Q^2 \ga
3GeV^2$  and negative and relatively large in absolute value at $Q^2=0$.
With $I_n(Q^2)$ the situation is similar. All this indicates  large
nonperturbative effects in $I(Q^2)$ at $Q^2 \la 1GeV^2$.

In \cite{29}  the model was suggested, which describes $I(Q^2)$ (and
$\Gamma(Q^2)$) at low and intermediate $Q^2$, where GDH sum rules and the
behaviour of $I(Q^2)$ at large $Q^2$ where fullfilled. The model had been
improved in \cite{30,28}.  (Another model with the same goal was
suggested by Soffer and Teryaev \cite{31}).

Since it is known, that at small $Q^2$ the contribution of resonances to
$I(Q^2)$  is of importance, it is convenient to represent $I(Q^2)$  as a sum
of two terms

\be
I(Q^2)  = I^{res}(Q^2) + I^{\prime}(Q^2),
\ee
where $I^{res}(Q^2)$ is the contribution of baryonic resonances.
$I^{res}(Q^2)$ can be calculated from the data on electroproduction of
resonances. Such calculation was done with the account of resonances up to
the mass $W = 1.8 GeV$ \cite{27}.

In order to construct the model for nonresonant part $I^{\prime}(Q^2)$
consider the analytical properties of $I(q^2)$ in $q^2$. As is clear from
(40),(41)  $I(Q^2)$ is the moment of the structure function, i.e. it is a
vertex function  with two legs, corresponding to ingoing and outgoing
photons and one leg with zero momentum. The most convenient way to
study of analytical properties of $I(q^2)$ is to consider a more general
vertex function $I(q^2_1, q^2_2; p^2)$, where the momenta of the photons are
different, and go to the limit $p \to 0, ~q^2_1 \to q^2_2 = q^2$. ~
$I(q^2_1, q^2_2; p^2)$ can be represented by the double dispersion relation:

$$ I(q^2) = \lim_{q^2_1 \to q^2_2=q^2,p^2\to 0}I(q^2_1,q^2_2;p^2) = \left \{
\int~ds_2 \int~ ds_1 ~\frac{\rho(s_1,s_2;p^2)}{(s_1-q^2_1)(s_2-q^2_2)}
+\right.$$

\be
\left. + P(q^2_1)~\int \frac{\varphi(s, p^2)}{s-q^2_2}ds + P(q^2_2)~\int
\frac{\varphi(s, p^2)}{s-q^2_1}ds \right \}_{q^2_1=q^2_2=q^2,p^2 \to 0}
\ee
The last two terms in (46)   are the substruction terms in the  double
dispersion relation, $P(q^2)$ is the polynomial. Since, according to (41),
$I(q^2)$ decreases at $q^2 \to \infty,~P(q^2)$  reduces to a constant,
$P(q^2)=Const$ and the constant subtraction term in (46) is absent. We are
interesting in $I(Q^2)$ dependence in the domain $Q^2 \la 1 GeV^2$. Since
after performed subtraction, the integrals  in (46)  are well converging,
one may assume, that at \\
$Q^2 \la 2 - 3 GeV^2$ the main contribution comes
from vector meson intermediate staties.  $\varphi$-meson  weakly
interacts with nucleon, so the general form of $I^{\prime}(Q^2)$ is

\be
I^{\prime}(Q^2) = \frac{A}{(Q^2 + \mu^2)^2} + \frac{B}{Q^2 + \mu^2},
\ee
where $A$ and $B$ are constants, $\mu$ is $\rho$  (or $\omega$) mass. The
constant $A$ and $B$  are determined from GDH sum rules at $Q^2=0$  and from
the requirement that at high $Q^2 \gg \mu^2$  takes place the relation

\be
I(Q^2) \approx I^{\prime}(Q^2) \approx \frac{2m^2}{Q^2}\Gamma^{as} (Q^2),
\ee
where $\Gamma^{as}(Q^2)$ is given by (5). $(I^{res}(Q^2)$  fastly decreases
with $Q^2$ and is very small above $Q^2=3 GeV^2$). These conditions are
sufficient to determine in unique way the constant $A$ and $B$ in (47). For
$I^{\prime}(Q^2)$ it follows:

\be
I^{\prime}(Q^2) = 2m^2 \Gamma^{as}(Q^2_0)\Biggl [ \frac{1}{Q^2 + \mu^2} -
\frac{c\mu^2}{(Q^2 + \mu^2)^2}\Biggr ],
\ee

\be
c = 1 + \frac{\mu^2}{2m^2}~\frac{1}{\Gamma^{as}(Q^2_0)}\Biggl [
\frac{1}{4}\kappa^2 + I^{res}(0)\Biggr ],
\ee
where $I^{res}(0)$ are given by the left column in (39).

The model and eq.49  cannot be used at high $Q^2 \ga 5 GeV^2$: one cannot
believe, that at such $Q^2$  the saturation of the dispersion relation (46)
by the lowest vector meson is a good approximation. For this reason there is
no matching of (49)  with QCD sum rule calculations of twist-4 terms.
(Formally, from (49) it would follow $b_{p-n} \approx -0.15,~ b_{p+n} \approx
-0.07$). It is not certain, what value of the matching point $Q^2_0$ should
be chosen in (49). This results in 10\% uncertainty in the theoretical
predictions. Fig.2 shows the predictions of the model in
comparison with recent SLAC data \cite{32}, obtained at low $Q^2 = 0.5$  and
$1.2 GeV^2$  as well as SMC and SLAC data at higher $Q^2$. The chosen
parameters are $\Gamma_p^{as}(Q^2_0) = 0.142, ~\Gamma_n^{as}(Q^2_0) =
-0.061$, corresponding to $c_p = 0.458, ~c_n = 0.527$  in (49),(50). The
agreement with the data, particularly at low $Q^2$, is very good. The change
of the parameters only weakly influences $\Gamma_{p,n}(Q^2)$ at low $Q^2$.
(For example, if we use instead of $\Gamma^{as}_p,~\Gamma^{as}_n,$
mentioned above, the values $\Gamma_p(Q^2_0) = 0.130, ~\Gamma_n(Q^2_0) =
-0.045$, then $\Gamma_p(Q^2)$ at $Q^2 < 1.5 GeV^2$ is shifted by less than
10\% and the values of $\Gamma_n(Q^2)$ by less than 20\%).

\vspace{5mm}
{\large \bf ~Conclusion.}

\vspace{3mm}
The nonperturbative $Q^2$-dependence of the sum rules for spin dependent
$e(\mu)N$ scattering is discussed. Two domains of $Q^2$  are considered. At
high $Q^2$  the determination of twist-4 corrections by QCD sum rule
approach  is analysed. It was found, that for the Bjorken sum rule the
twist-4 correction is small and, although its uncertainty is large, about
50\%, it does influences too  much Bjorken sum rule and the value of
$\alpha_s$ determined from this sum rule. However, in order to have the
reliable determination of $\alpha_s$  from comparison of the Bjorken sum
rule with the data, the accuracy of the latter must be improved by a factor
of 3-4.  For the singlet sum rule -- $p+n$  or proton separately -- the QCD
sum rule approach gives only the order of magnitude of twist-4 correction.
At $Q^2=5 GeV^2$ the twist-4 correction in this case is smaller than the
today experimental error. At low and intermediate  $Q^2$ a model was
presented, which realizes a smooth connection of GDH sum rules at $Q^2=0$
with the sum rules for $\Gamma_p(Q^2), ~\Gamma_n(Q^2)$  at high $Q^2$. The
agreement of the model with recent data is perfect.

\newpage
{\large \bf Acknowledgements}

\vspace{3mm}
I am thankful to V.Burkert for the information about the E143  data in the
resonance region and for the curves of resonance contributions which he
kindly presented me. This work was supported in part by the Russian
Foundation of Fundamental Research Grant 97-02-16131, US Civilian Research
and Development Foundation Grant RP2-132, INTAS Grant 93-0273 and
Schweizerische Nationalfond Grant 7SUPJO48716.

\newpage

\centerline{\large \bf Figure Captions}

\vspace{10mm}
\begin{tabular}{lp{12cm}}
{\bf Fig.~1} & The $Q^2$-dependence of integrals $I_p(Q^2), I_n(Q^2),
I_p(Q^2) - I_n(Q^2)$. The vertical axis is broken at negative values.\\
& \\
{\bf Fig.~2} & The $Q^2$-dependence of $\Gamma_p = \Gamma^{\prime}_p +
\Gamma^{res}_p$ (solid line), described by eqs.(45,49,50). $\Gamma^{res.}_p$
(dash-dotted) and $\Gamma^{\prime}_p$ (dashed) are the resonance and
nonresonance parts. The experimental points are: the dots from E143 (SLAC)
\cite{32}, the square - from E143 (SLAC) \cite{33}, the cross - SMC-SLAC
combined data \cite{1}, the triangle from SMC cite{1}.\\
& \\
{\bf Fig.~3} & The same
as in Fig.2 but for neutron. The experimental points are: the dots from E143
(SLAC) measurements on deuteron \cite{32}, the square at $Q^2 = 2 GeV^2$ is
the E142(SLAC) \cite{34} data from measurements on polarized ~$^3He$, the
square at $Q^2 = 3 GeV^2$ is E143(SLAC) \cite{35} deuteron data, the cross is
SMC-SLAC combined data \cite{1}, the triangle is SMC deuteron data \cite{1}.
\end{tabular}


\newpage

\begin{thebibliography}{99}
\bibitem{1}SMC-Collaboration, preprint CERN-PPE/96, submitted to Physical
Review D;

R.Voss, talk at the Workshop on High Energy Polarization Phenomena,

Ringberg, Germany, February 1997.
\bibitem{2} K.Griffioen for E143, E154, E155 Collaborations (SLAC),

talk at the Workshop on High Energy Polarization Phenomena,

Ringberg, Germany, February 1997.
\bibitem{3} S.Gerasimov, Yad.Fiz. {\bf 2} (1996) 930.
\bibitem{4} S.D.Drell and A.C.Hearn, Phys.Rev.Lett. {\bf 16} (1966) 908.
\bibitem{5} B.L.Ioffe, V.A.Khoze and L.N.Lipatov, Hard Processes
v.1, North Holland,

Amsterdam, 1984.
\bibitem{6} R.M.Barnett at al, Particle Data Group, Phys.Rev. {\bf D54}
(1996) 1.  \bibitem{7} J.Kadaira et al., Phys.Rev. {\bf D20} (1979) 627;

Nucl.Phys. {\bf B159} (1979) 99, {\bf 165} (1980) 129.
\bibitem{8} S.A.Larin and J.A.M.Vermaseren, Phys.Lett. {\bf 259} (1991) 345.
\bibitem{9} S.A.Larin, Phys.Lett. {\bf 334} (1994) 192.
\bibitem{10} S.A.Larin, T.van Ritbergen and J.A.M.Vermaseren,
preprint UM-TH-97-02,

NIKHEF-97-011 (1997).
\bibitem{11} A.L.Kataev, Phys.Rev. {\bf D50} (1994) 5469.
\bibitem{12} S.Y.Hsueh et al., Phys.Rev. {\bf D38} (1988) 2056.
\bibitem{13} R.D.Carlitz, J.C.Collins and A.H.Mueller, Phys.Lett. {\bf B
214} (1988) 229.
\bibitem{14} S.D.Bass, B.L.Ioffe, N.N.Nikolaev and A.W.Thomas,
J.Moscow Phys.Soc. {\bf 1}

(1991) 317.
\bibitem{15} B.L.Ioffe in: International School of Nucleon Structure,
1-st Course:

The Spin Structure of the Nucleon, Erice-Sicily, August 1995,

Ed.'s B.Frois, V.Hughes, Plenum, in press.
\bibitem{16} E.V.Shuryak and A.I.Vainshtein, Nucl.Phys. {\bf 201} (1982)
144.
\bibitem{17} I.I.Balitsky, V.M.Braun and A.V.Kolesnichenko,
Phys.Lett. {\bf 242} (1990) 245,

Errata {\bf B318} (1993) 648.
\bibitem {18} B.L.Ioffe, Nucl.Phys. {\bf B188} (1981) 317.
\bibitem{19} B.L.Ioffe and A.V.Smilga, Nucl.Phys. {\bf B232} (1984) 109.
\bibitem{20} B.L.Ioffe , Phys.At.Nucl. {\bf 58} (1995) 1408.
\bibitem{21} A.Oganesian, preprint ITEP, in preparation.
\bibitem{22} B.L.Ioffe and A.Yu.Khodjamirian, Yad.Fiz. {\bf 55} (1992) 3045.
\bibitem{23} M.Gl\"uck and E.Reya, Phys.Lett.{\bf B270} (1991) 65.
\bibitem{24} R.L.Heiman, Nucl.Phys. {\bf B64} (1973) 429.
\bibitem{25} I.Karliner, Phys.Rev. {\bf D7} (1973) 2717.
\bibitem{26} R.L.Workman and R.A.Arndt, Phys.Rev. {\bf D45} (1992) 1789.
\bibitem{27} V.Burkert and Z.Li, Phys.Rev. {\bf 47} (1993) 46.
\bibitem{28} V.Burkert and B.L.Ioffe, JETP, {\bf 105} (1994) 1153.
\bibitem{29} M.Anselmino, B.L.Ioffe and E.Leader, Sov.J.Nucl.Phys. {\bf 49}
(1989) 136.
\bibitem{30} V.Burkert and B.L.Ioffe, Phys.Lett. {\bf B296} (1992) 223.
\bibitem{31} J.Soffer and O.Teryaev, Phys.Rev. {\bf D51} (1995) 25.
\bibitem{32} K.Abe et al., (SLAC E143 Collaboration) Phys.Rev.Lett.
{\bf 78} (1977)  815.
\bibitem{33} K.Abe et al. (SLAC E143 Collaboration),
Phys.Rev.Lett. {\bf 74} (1995) 346.
\bibitem{34} P.L.Anthony et al. (SLAC E142 Collaboration),
Phys.Rev. {\bf D54} (1996) 6620.
\bibitem{35} K.Abe et al. (SLAC E143 Collaboration),
Phys.Rev.Lett. {\bf 75} (1995) 25.
\end{thebibliography}
\end{document}